\documentclass[useAMS,usenatbib]{mnras}
\usepackage{graphicx}
\usepackage{float}
\usepackage{caption}
\usepackage{subcaption}
\usepackage[authoryear]{natbib}
\usepackage{color}
\usepackage{amsmath}
\usepackage{soul,ulem}
\title[A 3D MHD study of the SN\,1006: the turbulent case ]{A 3D MHD
  simulation of SN\,1006: a polarized emission study for the
  turbulent case}
\author[Vel\'azquez et al.]{P. F. Vel\'azquez$^{1}$\thanks{Email:
    pablo@nucleares.unam.mx}, E. M. Schneiter$^{2,3}$, E. M. Reynoso$^{4}$, A. Esquivel$^{1}$, F. De Colle$^{1}$,  \and J. C. Toledo-Roy$^{5}$, D.O. G\'omez $^{4,6}$, M.V. Sieyra$^{2}$, A. Moranchel-Basurto$^{7}$ 
  \\ $^{1}$Instituto de
  Ciencias Nucleares, Universidad Nacional Aut\'onoma de M\'exico, Ciudad de 
M\'exico,  M\'exico\\ $^2$Instituto de Astronom\'\i a Te\'orica y Experimental, Universidad
  Nacional de C\'ordoba, C\'ordoba, Argentina\\ $^{3}$Departamento de
  Materiales y Tecnologia, UNC, C\'ordoba, Argentina \\ 
   $^{4}$CONICET-Universidad de Buenos Aires. Instituto de Astronom\'\i a y F\'\i sica del
  Espacio, Suc. 28, CP: 1428, Buenos Aires, Argentina\\ 
  $^5$Centro de Ciencias de la Complejidad, Universidad
Nacional Aut\'onoma de M\'exico, Ciudad de M\'exico, M\'exico.\\
  $^6$ Departamento de F\'\i sica, Facultad de Ciencias Exactas y Naturales, Universidad de Buenos Aires, Buenos Aires, Argentina\\
  $^7$Escuela Superior de F\'\i sica y Matem\'atica, Instituto Polit\'ecnico Nacional, Ciudad de M\'exico, M\'exico.}
\begin{document}

\date{Accepted  . Received }

\pagerange{\pageref{firstpage}--\pageref{lastpage}} \pubyear{2002}

\maketitle

\label{firstpage}

\begin{abstract}
Three dimensional magnetohydrodynamical simulations were carried out in order to
perform a new polarization study of the radio emission of the supernova remnant
SN 1006.  These simulations consider that the remnant expands into a
turbulent interstellar medium (including both magnetic field and
density perturbations). Based on the referenced-polar angle technique,
a statistical study was done on observational and numerical magnetic
field position-angle distributions. Our results show that a turbulent medium with an adiabatic index of 1.3 can reproduce the polarization properties of the SN 1006 remnant. This statistical study reveals itself as
a useful tool for obtaining the orientation of the ambient 
magnetic field, previous to be swept up by the main supernova remnant shock.

\end{abstract}


\section{Introduction}

SN 1006 is a young supernova remnant (SNR) that
has gained interest due to the large observational data
available in a wide wavelength range, therefore
serving as an ideal laboratory when trying to understand the
observed morphology in each frequency.
This remnant is classified as part of the bilateral SNR group,
whose main characteristic is the presence of
two bright and opposite arcs in radio-frequencies.

The morphology and emission of this type of SNRs is largely
determined by the environment in which they evolve, sometimes having an
irregular or clumpy background and/or with the presence of density and
magnetic fields gradients.

The interstellar medium is known to be turbulent, as inferred from observational data \citep{lee1976} and local information on the interplanetary medium \citep{jokipii1969}. Such  turbulent medium, with a Kolmogorov-like power spectrum that spans over a large range of spatial scales \citep{guo2012}, gives rise to Rayleigh-Taylor (RT) instabilities during the fast expansion of the SNR, which in turn affects the radio emission and is imprinted in the measured Stokes parameters.

\citet{jun1999} performed 2D MHD simulations of the evolution of a
young Type Ia SNR interacting with an
interstellar cloud. They found that the interaction produced RT
instabilities causing amplification of the
magnetic field, and subsequently
changing the synchrotron
brightness. This work emphasized the importance of including a turbulent medium when trying
to match observations.  In this direction, \citet{balsara2001a} performed
more realistic 3D simulations of a SNR evolving in a turbulent
background. 
They found that the  variability of the emission during the early phase
of SNR evolution gives information about the turbulent
environment. At the same time, that the interaction with such a turbulent environment  generates an enhanced turbulent post-shock region 
which has observational consequences, as it affects the acceleration of 
relativistic particles.

\citet{guo2012} employed 2D MHD simulations to
  study the magnetic field amplification mechanisms during the propagation of the SNR blast wave.
  They found that, for high resolution simulations,
  magnetic field growth at small scales occurs
  efficiently in two distinct
  regions: one related to the shock amplification, and the 
  other, 
  to the RT-instabilities at the contact discontinuity between
  the shocked ejecta and the shocked interstellar medium.

\citet{fang2012} carried out a study similar to that of \citet{guo2012}
but considering different adiabatic indexes
to account for both the diffusive shock acceleration and the
escape of the accelerated particles from the shock. They found that a
smaller effective adiabatic index 
produces a larger magnetic field enhancement.  They
extended their study considering a scenario where the expanding
ejecta interacts with a large,  dense clump and found,
in addition to an increase in the complexity of the magnetic field, the
development of a non-thermal emitting filament.
%
In a subsequent study, \citet{fang2014} performed 3D  MHD simulations
considering a turbulent medium with a Kolmogorov-like power spectrum
with parameters resembling the SNR RX J0852.0-4622. Their X-ray and
$\gamma$-ray synthetic maps showed the expected rippled
morphology with a broken circular shape along the shell, nicely reproducing 
the observations.

In general, the diffusive shock acceleration is driven by magnetic fields strong enough to generate non-thermal
emission. Therefore, any structure or dynamic interaction that results
in the amplification of the magnetic field has important consequences
on the observed brightness.

According to the non-thermal radio-emission morphology, SN 1006
has been classified as a bilateral or {\it barrel-shaped} SNR
\citep{kesteven1987}, showing two main bright arcs towards the NE and
SW.  The non-thermal X-ray emission is also predominant in the NE and
SW rims \citep{cassam2008}.

SN 1006 is located at a high Galactic latitude and
is therefore thought to be evolving in a fairly
homogenous interstellar medium.  In \citet{schneiter2010} and
\citet{schneiter2015} we made use of this assumption and presented
studies of the system in 2D and 3D, respectively. In the
latter work we further introduced synthetic maps of the Stokes $Q$ and
$U$ parameters wich allowed a better and more direct comparison with
the observations \citep{reynoso2013}.

Recently, \citet[a]{west2016a} \citep[see
    also,][b]{west2016b} studied radio emission for a sample of
  Galactic axisymmetric SNRs, including SN 1006. They concluded
  that the quasi-perpendicular acceleration mechanism can sucessfully
  fit the observed morphology of the remnants in their sample, although SN 1006 seems to be the exception, since the quasi-parallel
  mechanism agrees better with observations. 

 In the present work we use the same SNR initialization as in
the two previous works of \citet{schneiter2010} and
\citet{schneiter2015} but relaxing the assumption of a homogeneous
ISM by considering that the SNR evolves into a turbulent medium as it
was shown in \citet{yu2015} \citep[see also][]{fang2014}.

The goal of this work is to
propose an alternative method to determine the
position angle of the ambient
magnetic field, based on the polar-referenced angle method employed by
\citet{reynoso2013} and \citet{schneiter2015}.

The numerical model is presented in  section \S
\ref{numerical}. For completeness we recall the basic model employed by \citet{schneiter2015} in \S
\ref{model} and 
in section \S \ref{turbulence} we explain how the
turbulence is introduced. Section \S \ref{synthetic} explains how synthetic radio maps were obtained and the results are presented in section \S \ref{results}. The discussion and final
conclusions are given in section \S \ref{discussion}.

\section{Numerical model}\label{numerical}

\begin{figure}
        \centering
        \includegraphics[width=0.4\textwidth]{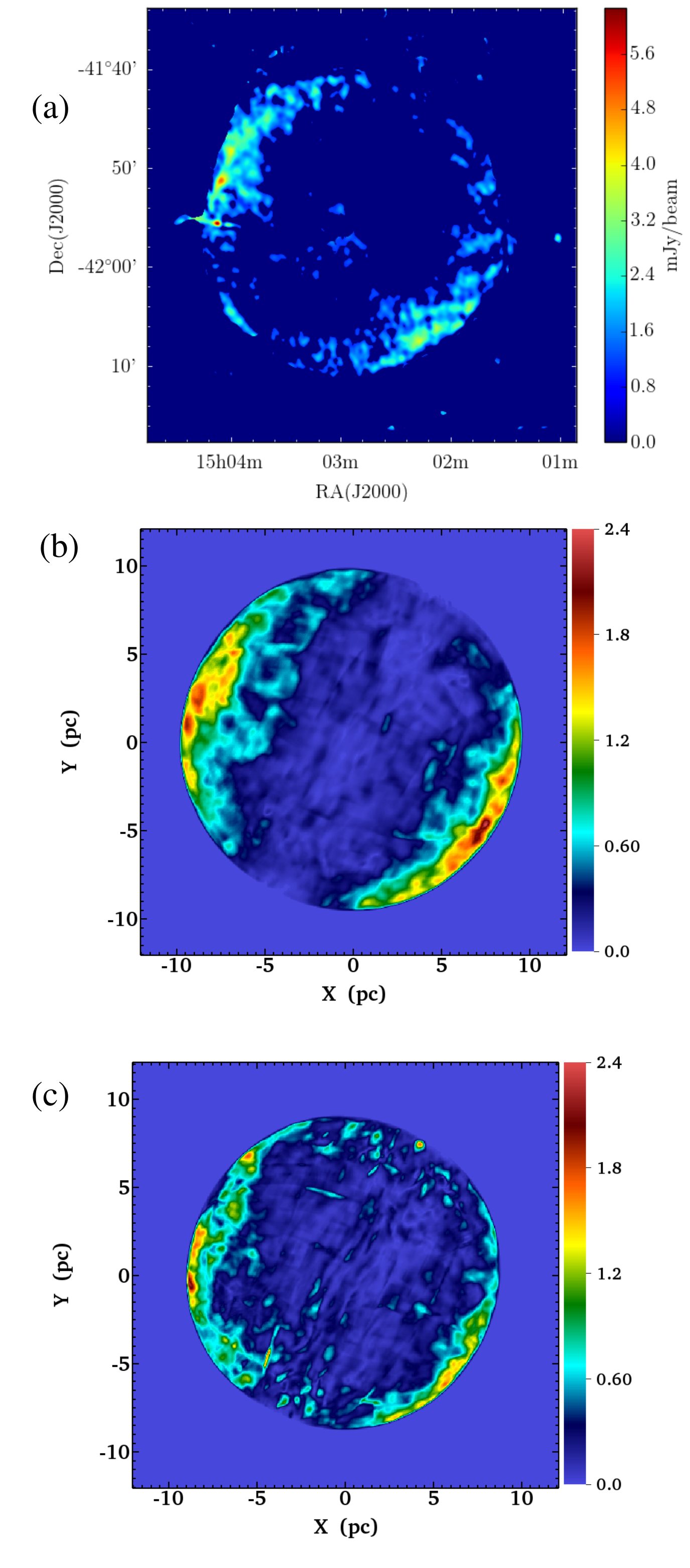}

        \caption{Comparison of the linearly polarized intensity at 1.4
          GHz between the observations (panel (a)) and synthetic maps
          of a SNR expanding into a turbulent medium with adiabatic index $\gamma=5/3$ (model R1b, panel (b)) and $\gamma=1.3$ (model R2b, panel(c))}
        \label{fig:sincro}
\end{figure}

The numerical model employed is basically the same as in
\citet{schneiter2010,schneiter2015}, with the additional
assumption of a turbulent background, where both the density and
magnetic fluctuations have a 3D Kolmogorov-like power
spectrum, as will be explained further down.

To simulate the evolution of SN 1006 we employed the \emph{Mezcal} code
\citep{decolle2006,decolle2008,decolle2012}. This code solves the full
set of ideal MHD equations in a Cartesian geometry with an adaptive
mesh, and includes a cooling function to account for radiative losses
\citep{decolle2006}.

The computational domain is a cube of $24$ pc per side, which
we will denote
as ($x,y,z$), and is discretized on a five level
binary grid with a maximum resolution of $4.7\times 10^{-2}$ pc. All
the outer boundaries were set to outflow condition.

\subsection{SNR initial conditions}
\label{model}
In what follows, we describe the physical parameters and setup used to simulate the
SNR, which are the same as those presented in \citet{schneiter2015}.

A supernova explosion is initialized by the deposition of $E_{0}=2.05\times 
10^{51}$ erg in a radius of $R_0=0.65$ pc
located at the centre of the computational domain. The energy is
distributed such that $95\%$ of it is kinetic and the remaining $5\%$ is
thermal. 

The ejected mass was distributed in two parts: an inner homogeneous
sphere of radius $r_c$ containing $4/7$ths of the total mass
(M$_*=1.4$M$_\odot$) with a density $\rho_c$, and an outer shell
containing the remaining $3/7$ths of the mass
following a power law ($\rho \propto r^{-7}$) as in \citet{jun1996a}.
The velocity has a increasing linear profile with $r$, which reaches a value of
$v_0$ at $r=R_0$. The parameters $\rho_c$, $r_c$, and $v_0$ are
functions of $E_0$, $M_*$, and $R_0$, and were computed using
equations (1)-(3) of \citet{jun1996a}.

\subsection{The turbulent background }
\label{turbulence}

\begin{figure}
        \centering
        \includegraphics[width=0.4\textwidth]{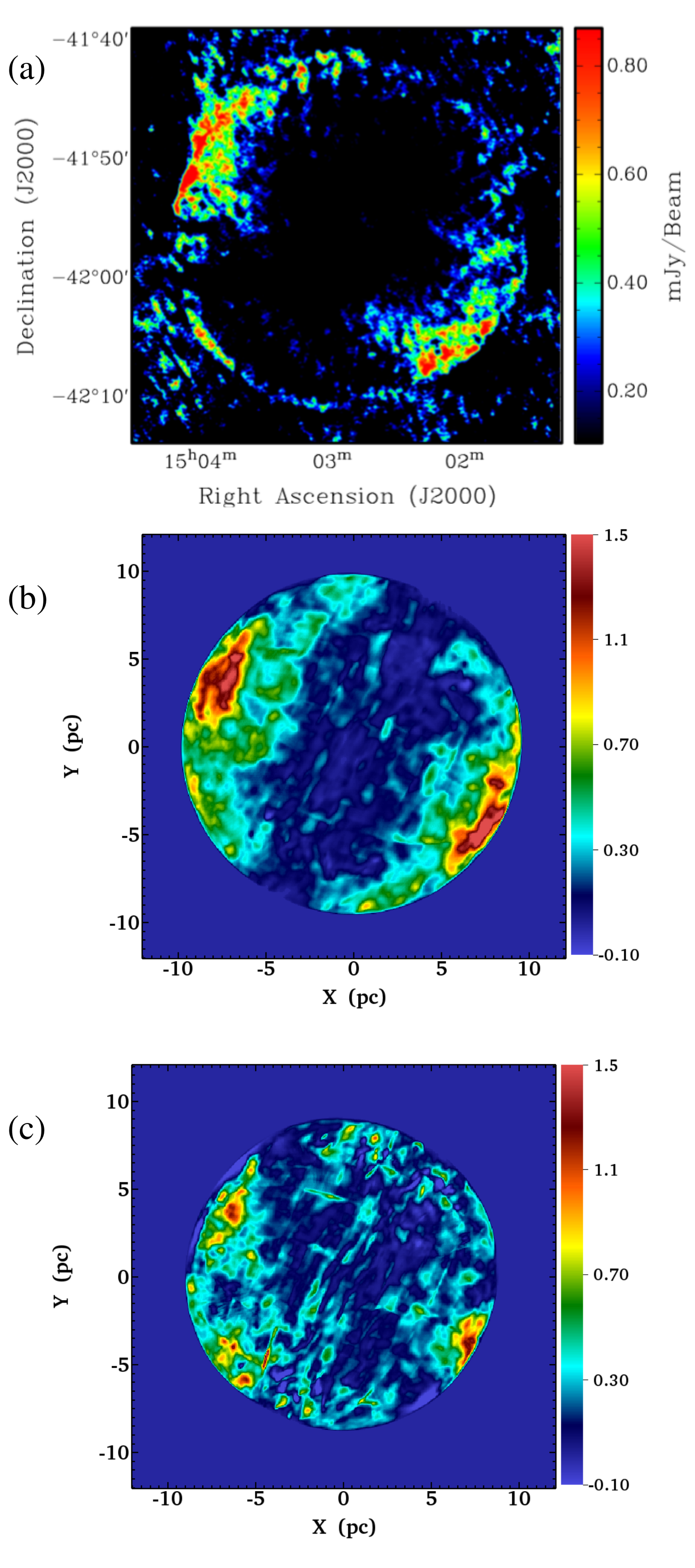}

        \caption{Same as Figure \ref{fig:sincro} but for maps of the
Stokes parameter Q.}\label{fig:fluxQ}
\end{figure}

\citet{jun1999} suggested that 
the turbulent structures of the brightest radio emission correlate well with the
magnetic field in general. For this reason we performed new simulations with
a more realistic magnetized turbulent interstellar medium.  
In what follows, we will make use of equations (10) to (15)
of \citet{yu2015}, which we reproduce below.
To introduce the turbulent
background we assumed a 3D Kolgomorov-like power
spectrum fluctuation for both the density and the magnetic
field, similar to that of \citet{jokipii1987} 
\citep[see also][]{yu2015,fang2014}. The power spectrum follows:

\begin{equation}
P \propto \frac{1}{1+(k L_c)^{11/3}}.
\end{equation}

To generate the turbulence we assume a coherence length of $L_c=3\,
pc$  and a wavenumber $k=\frac{2 \pi}{L}$.
Our simulations are computed with $N_m=903$ wave modes, and $L$ varies
between $L_{min}=\Delta x$ and $L_{max}=L_{sim}$, being $\Delta x$ and $L_{sim}$ the cell and the computational domain sizes, respectively.
%
%

\noindent The initial magnetic field is given by:

\begin{equation}
{\bf B}(x,y,z)={\bf B}_0+{\bf \delta B},
\end{equation}
where ${\bf B_0}= B_0(x) \hat{y}$ and the magnetic field perturbation
is: 
\begin{equation}
{\bf \delta B}=\Re\bigg[\sum_{n=1}^{N_m} A(k_n)\left( \cos \alpha_n \hat{x}' + i \sin \alpha_n  \hat{y}' \right) \exp (i k_n z_n')\bigg],
\end{equation}
with 

\begin{equation}
A^2(k_n)=\sigma_B^2 \frac{\Delta V_n}{1+(k_nL_c)^{11/3}}\sum_{n=1}^{N_m} \left [\frac{\Delta V_n}{1+(k_nL_c)^{11/3}} \right]^{-1},
\end{equation}
where $\sigma^2_B$ is the wave variance of the magnetic field, and
the normalization factor, $\Delta V_n$, is given by:

\begin{equation}
\Delta V_n=4 \pi k_n^2 \Delta k_n.
\end{equation}

Both systems $(x',y',z')$ and $(x,y,z)$ are related by: 
\[ 
\begin{pmatrix}
 x' \\
 y' \\
 z'
\end{pmatrix}
=
\begin{pmatrix}
\cos \theta_n \cos \phi_n& \cos \theta_n \sin \phi_n & -\sin \theta_n \\
-\sin \phi_n & \cos \phi_n &  0 \\
\sin \theta_n \cos \phi_n & \sin \theta_n \sin \phi_n & \cos \theta_n  
\end{pmatrix}
\begin{pmatrix}
 x \\
 y \\
 z
\end{pmatrix}
\] 
where $\theta_n$ and $\phi_n$
represent  the  direction  of  propagation  of  the  wave
mode $n$ with the wave number $k_n$ and polarization $\alpha_n$.

For the density fluctuation we employed the same log-normal
distribution as in \citet{giacalone2007}:

\begin{equation}
n(x,y,z)=n_0 \exp({f_0+\delta f}),
\end{equation}

\noindent where $f_0$ is a constant and $\delta f$ is the density perturbation with a wave variance
$\sigma^2_d$. Both $\delta f$ and $\bf{\delta B}$ have the same 3D
Kolmogorov-like power spectral index.

The turbulent environment temperature and number density were set to
T$_0=10^{4}$K and $n_0=5\times 10^{-2} $cm$^{-3}$, respectively. The
wave variances for magnetic field and density were set as
$\sigma^2_B=0.25\, B_0^2$ and $\sigma^2_d=0.4 n_0^2$.

As mentioned above, one of the purposes of this work is to be able
to study the effect of a perturbation on some observed
parameters. To compare with the observations, the synthetic maps were performed for an integration time corresponding to $1000$ yr  (the age of SN 1006).

\section{Synthetic emission maps}\label{synthetic}

\begin{figure}
  \centering
   \includegraphics[width=0.4\textwidth]{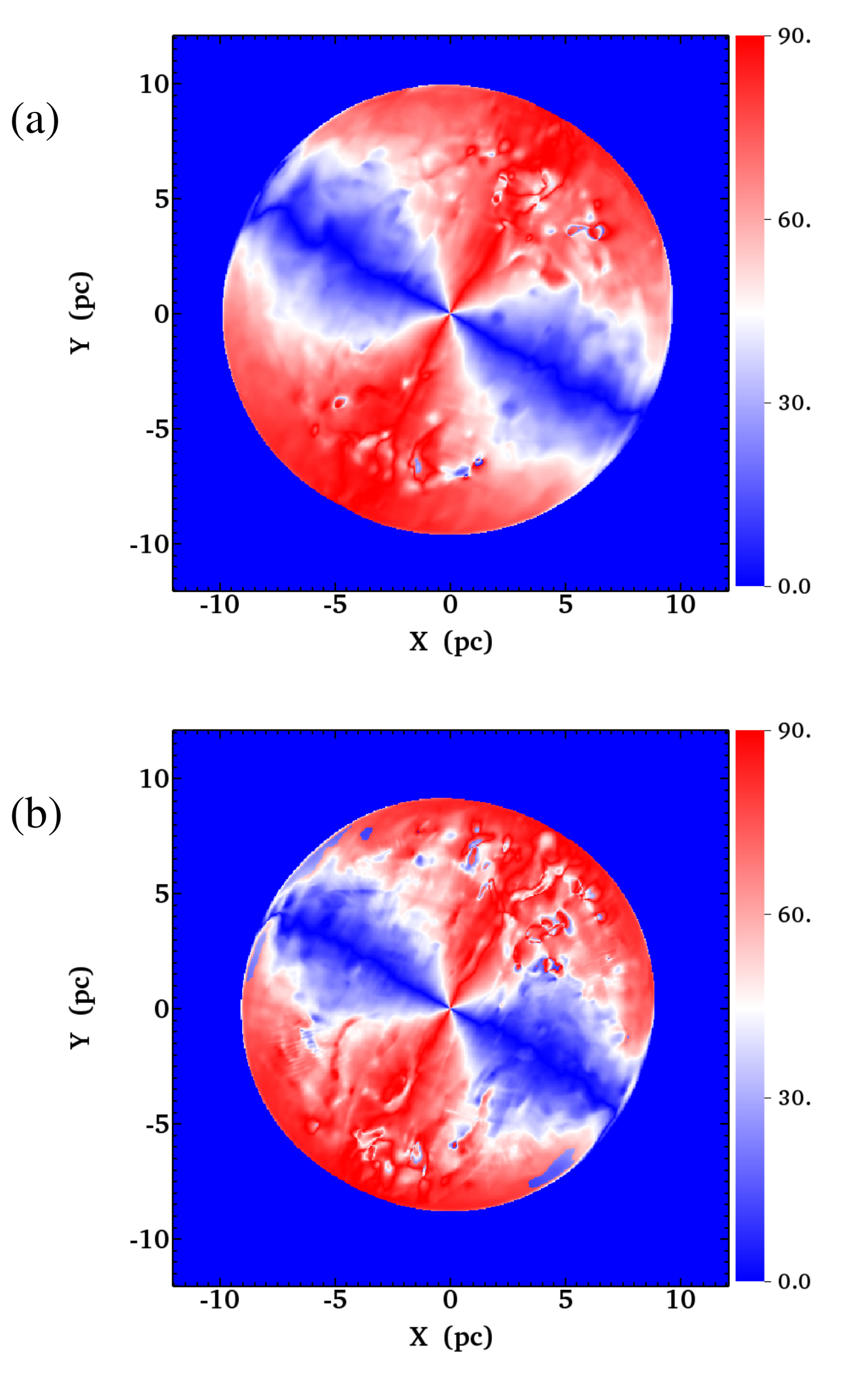}
    \caption{Synthetic polar-referenced angle maps obtained
      from numerical simulations of the SNR 1006 expanding into a turbulent medium and with adiabatic index $\gamma=5/3$ (model R1b, panel (a)) and $\gamma=1.3$ (model R2b, panel(b)) 
      The axes are in pc and the
      linear color scale is given in degrees.
    }
\label{refangles}
\end{figure}

\begin{table}
\begin{center}
  \begin{tabular}{cccc}
            \hline
            \noalign{\smallskip}
     run  &  $\gamma$ &  $\delta${\bf B} & $\delta\rho$\\
             \hline
           \noalign{\smallskip}
R1a  &  $5/3$  &  no & no\\
R1b  &  $5/3$  &  yes & yes\\
R2a  &  $1.3$ &  no & no\\
R2b  &  $1.3$ &  yes & yes\\
R2c  &  $1.3$ &  yes & no\\
R2d  &  $1.3$ &  no & yes\\
           \hline
            \noalign{\smallskip}
  \end{tabular}
  \caption{Runs carried out in this paper with their corresponding hypothesis: adiabatic index, turbulent magnetic field and/or turbulent density distribution.}
  \label{tabrun}
\end{center}
\end{table}

In Table 1 we list the runs that were carried
out. With these runs we want to
analyze the effects of both 
the reduction of the value of the adiabatic index $\gamma$
and/or 
the inclusion of a turbulent background with a
Kolgomorov-like power spectrum for both density and magnetic
field. These runs are labeled as R$i$p, where $i$ is 1 for the case
of simulations with an adiabatic index $\gamma=5/3$, and 2 for $\gamma=1.3$.
The label ``p'' indicates the presence or absence of turbulence in the background magnetic field and density: p = a indicates no turbulence in either variable; p = b includes turbulence in both; p = c only considers a turbulent $\bf B$; and p = d only considers a turbulent $\rho$.

Synthetic radio emission maps were obtained
from the numerical results. In order to compare them
with the observations, the computational domain (denoted as $xyz$ system), they were  rotated
with respect to the ``image'' system
($x_i y_i z_i$ system).  At the beginning both
systems coincide. The $x_iy_i$ plane of the ``image'' system is the plane of the sky,
being $\hat y_i$ the direction towards the ``North'', $-\hat x_i$ the direction 
towards the ``East'', and the line of sight (hereafter LoS) the $\hat z$
direction.  Then, synthetic maps were obtained after
performing three rotations on the computational
domain. First, a rotation around $\hat{y}_i$ was applied (the
``North'')  with an angle
$\varphi_{y_i}$. Second, a rotation in $\varphi_{x_i}$ was carried out
around the $\hat{x}_i$-direction. Finally, the computational domain was rotated by $\varphi_{z_i}$
around the $\hat{z}_i$-direction (the line of sight). To get the best agreement with the observations, after  several tests
these angles were set as $-15\degr$, $-30\degr$, and $60\degr$ for
$\varphi_{x_i}$, $\varphi_{y_i}$, and $\varphi_{z_i}$, respectively  . In this way,
the projection of the unperturbed magnetic field ${\bf B_0}$ was tilted
by $60\degr$ with respect to the $\hat{y}_i$ direction.

\subsection{Synchrotron emissivity}

For each point $(x_i,y_i,z_i)$ of the SNR we can obtain
the synchrotron specific intensity as \citep[see][and references
  therein]{cecere2016}:

\begin{equation}
j(x_i,y_i,z_i,\nu) \propto K \rho\ v^{4\alpha}\ B_{\perp}^{\alpha +1} \nu^{-\alpha},
\end{equation}
where $\nu$ is the observed frequency, $B_{\perp}$ is the component of
the magnetic field perpendicular to the LoS, $\alpha$ is the 
spectral index -which was set to $0.6$ for this object- and $\rho$ and
$v$ are the density and velocity of the gas, respectively. The
coefficient $K$ includes the obliquity dependence, being either
proportional to $\sin^2{\Theta_{Bs}}$ for the quasi-perpendicular case or to 
$\cos^2{\Theta_{Bs}}$ for the quasi-parallel case. The angle $\Theta_{Bs}$ is
the angle between the shock normal and the post-shock magnetic field
\citep{fulbright1990}. For the case of SN 1006,  \citet{bocchino2011}
and \citet{reynoso2013} showed that the observed morphology of this
remnant can be explained by considering the quasi-parallel
case. \citet{schneiter2015} carried out a polarisation study of this
remnant, based on
3D MHD simulations, finding that only the quasi-parallel case can succesfully
reproduced the observed morpholoy of the Stokes parameter Q
\citep[see Figure 3 of][]{schneiter2015}.
For these reasons only the quasi-parallel case was considered in this
work.
The synthetic synchrotron emission maps are obtained by integrating
$j(x_i,y_i,z_i,\nu)$ along the LoS or $z_i$-axis, i.e.:

\begin{equation}
I(x_i,y_i,\nu)=\int_{\mathrm{LoS}} j(x_i,y_i,z_i,\nu) dz_i ,
\label{sincromap}
\end{equation}
being $x_i$ and $y_i$ the coordinates in the plane of the sky.

\subsection{Stokes parameters and position angle distribution maps}
%
%
\begin{figure}
  \centering
   \includegraphics[width=0.5\textwidth]{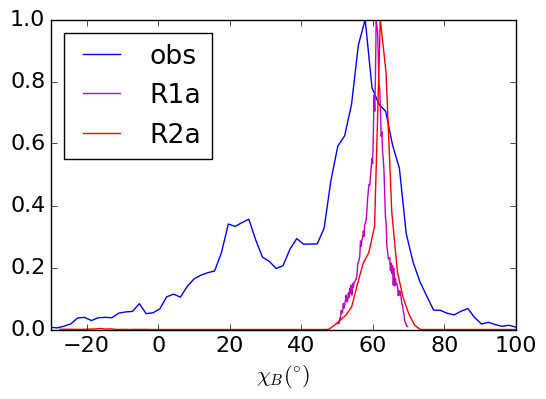}
    \caption{Comparison of the normalized distributions (with respect to their maxima) of the magnetic field
      position-angle 
      as 
      obtained from the observations and from
      the runs R1a and R2a (different $\gamma$ and both
       corresponding to a SNR propagating into a uniform medium, i.e. with   $\delta${\bf B} = $\delta\rho=0$; see table \ref{tabrun}), applying the
      polar-referenced angle selection.}%
\label{compnonoise}
\end{figure}

With the purpose of comparing the numerical results with the
observations, synthetic maps of the Stokes
parameters $Q$ and $U$  were calculated as follows
  \citep[][]{clarke1989,jun1996b,schneiter2015}:

\begin{equation}
Q(x_i,y_i,\nu)=\int_{\mathrm{LoS}} f_0  j(x_i,y_i, z_i,\nu) \cos\left[ 2\phi(x_i,y_i,z_i)\right] dz_i ,
\label{factorQ}
\end{equation}
\begin{equation}
U(x_i,y_i,\nu)=\int_{\mathrm{LoS}} f_0  j(x_i,y_i, z_i,\nu) \sin\left[ 2\phi(x_i,y_i,z_i)\right] dz_i ,
\label{factorU}
\end{equation}
where $\phi(x_i,y_i,z_i)$ is the position angle of the local electric field
in the plane of the sky  and $f_0$ is the degree of
linear polarization, which is a function of the spectral index $\alpha$:
\begin{equation}
f_0=\frac{\alpha +1}{\alpha + 5/3} .
\end{equation}
The position angle of the local magnetic field $\phi_{\mathrm{B}}(x_i,y_i,z_i)$ is known from the simulations, and the position angle of the local electric field $\phi(x_i,y_i,z_i)$ can be obtained from
$\phi_{\mathrm{B}}(x_i,y_i,z_i)$
by applying a $\frac{\pi}{2}$ rotation and correcting for Faraday rotation, as:

\begin{equation}
\phi(x_i,y_i,z_i)=\phi_{\mathrm{B}}(x_i,y_i,z_i)-\frac{\pi}{2}+\Delta\chi_{\mathrm{F}} .
\label{angles}
\end{equation}

\noindent The Faraday correction term $\Delta\chi_{\mathrm{F}}$ is  given by:
\begin{equation}
\Delta\chi_{\mathrm F}=\mathrm{RM}\ \lambda^2 ,
\label{rotfaraday}
 \end{equation}
where RM is the rotation measure (in units of $\mathrm{rad\ m^{-2}}$) and $\lambda$ is the wavelength of the observations (given in m).
  In the present study, only the external RM arising in the foreground ISM is considered. \citet{bandiera2016}  explored the effects of internal RM on the polarized synchrotron emission of SNRs. For the case of SN 1006, we have carried out a rough estimation of the internal RM, obtaining a value close to 10\% of the reported one by \citet{reynoso2013}  who obtained a RM of  12~$\mathrm{rad\ m^{-2}}$ for this remnant. 

The linearly polarized intensity is computed as: 

\begin{equation}
I_P(x_i,y_i,\nu)= \sqrt{Q(x_i,y_i,\nu)^2+U(x_i,y_i,\nu)^2}
\label{ipol}
\end{equation}
and the maps of the polarization angle distribution were computed as follows:

\begin{equation}
\chi(x_i,y_i)=\frac{1}{2}\tan^{-1}(U(x_i,y_i,\nu)/Q(x_i,y_i,\nu))
\label{chipol}
\end{equation}

Similarly, the distribution of the magnetic field orientation can be
calculated as:

\begin{equation}
\chi_{\mathrm B}(x_i,y_i)=\frac{1}{2}\tan^{-1}(U_{\mathrm B}(x_i,y_i,\nu)/Q_{\mathrm B}(x_i,y_i,\nu))
\label{chipolB}
\end{equation}
where $U_{\mathrm B}(x_i,y_i,\nu)$ and $Q_{\mathrm B}(x_i,y_i,\nu))$ are
calculated with the equations (\ref{factorQ}) and (\ref{factorU})
 and   replacing
$\phi(x_i,y_i,\nu)$  with
$\phi_{\mathrm B}(x_i,y_i,\nu)$.

\section{Results}\label{results}



\subsection{Synthetic polarization maps}

\begin{figure}
  \centering
   \includegraphics[width=0.5\textwidth]{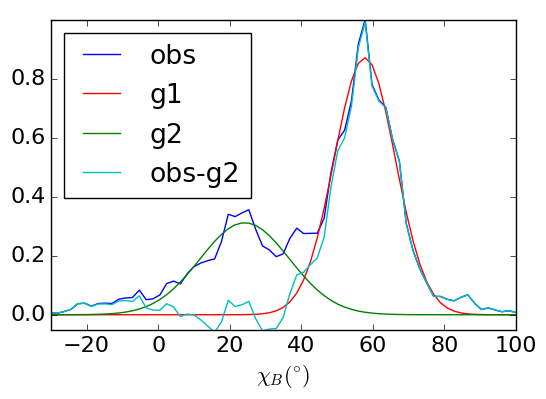}
    \caption{Two Gaussian fit of the observed position-angle
      distribution.}%
\label{gaussobs}
\end{figure}

Figure \ref{fig:sincro} shows a comparison between the observed map of
the linearly polarized intensity of SN 1006 (upper panel), and the
synthetic polarization maps obtained for runs with $\gamma=5/3$ (run
R1b, middle panel) and $\gamma=1.3$ (run R2b, bottom panel) by
using Equation (\ref{ipol}). The observed map was constructed by combining data 
obtained with the ATCA and the VLA at 1.4 GHz, as explained in \citet{reynoso2013}. The polarized emission image, as well as the distribution of the Q and U parameters, were convolved to a $15^{\prime\prime}$ beam. The two models used to construct the synthetic maps consider that the SNR is expanding into a turbulent medium with perturbations in both density and magnetic field. These synthetic
maps are shown in arbitrary units. Two opposite bright, noisy
arcs are observed in both synthetic maps, having a striking resemblance to the observations.  Reducing the value of $\gamma$
  produces a 5\% smaller expansion radius of the remnant as measured on the linearly polarized intensity maps.

Following the analysis developed by \citet{schneiter2015}, we compare the observational and synthetic maps of the Stokes parameter Q, obtained
for the quasi-parallel case using Equation
\ref{factorQ}. These maps are shown in Figure
\ref{fig:fluxQ}. A good overall
agreement between observations and numerical results is obtained, since
the opposite bright arcs are well reproduced. Also, note that these arcs are slightly more elongated than the ones obtained for
  the non-turbulent ISM \citep{schneiter2015}.

\subsection{Polar-referenced angle and a statistical study}

\begin{figure}
  \centering
   \includegraphics[width=0.5\textwidth]{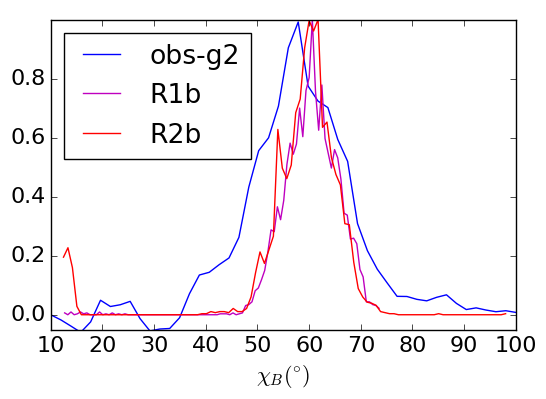}
    \caption{Same as Figure \ref{compnonoise} but comparing the distribution obtained for observations (excluding the contribution of
      Gaussian g2) and runs R1b and R2b (with different
      $\gamma$ and  for a turbulent medium, i.e.
      including both $\delta\rho$ and $\delta${\bf B}). }%
\label{compnoise}
\end{figure}

\begin{table}
\begin{center}
  \begin{tabular}{lcc}
            \hline
            \noalign{\smallskip}
     Gaussian   &  mean($\degr$) &  $\sigma$($\degr$)\\
             \hline
           \noalign{\smallskip}
g1  &  $58.0$  &  $9.0$ \\
g2  &  $23.0$  &  $13.0$ \\
\hline
            \noalign{\smallskip}
  \end{tabular}
\caption{Two Gaussian fit of the observational position-angle distribution}
\end{center}

\label{tab2}
\end{table}

The polar-referenced angle distribution $\chi_r$ is given by:
\begin{equation}
 \chi_r=\cos^{-1}(\hat r \cdot {\hat b}_{\perp}),
\label{chir}
\end{equation}
where $\hat{r}=(-x,y)/\sqrt{x^2+y^2}$  is the radial direction and
$\hat{b}_{\perp}=(-sin(\chi_{\mathrm B}), cos(\chi_{\mathrm B}))$ is  the
direction of the magnetic field perpendicular to the LoS, which is
obtained through Equation (\ref{chipolB}).

By using Equation (\ref{chir}), polar-referenced angle maps are
obtained for runs R1b and R2b, which are shown in Figure
\ref{refangles}. The blue color indicates
the region where the magnetic field is almost  parallel to the
radial direction, while red color highlights the 
regions where it is mostly perpendicular.

Given that the polar-referenced angle technique is a
  useful tool for estimating the direction of the pre-shock ISM magnetic field
  \citep[see][]{reynoso2013,schneiter2015} we carried out a statistical study
  based on the analysis of the position-angle distribution in those
  regions where $\chi_r$ approaches zero.

 The analysis was performed for both the numerical results and the
 observations in regions with synchrotron intensities larger than
 $10$\% of the maximum, and where the condition $\chi_r\le 14\degr$ (the observational
 angle error) is satisfied. These criteria were chosen to assure a good signal to noise ratio of the sample. The results are shown in Figure
 \ref{compnonoise}.  To obtain the observational distribution, electric vectors were computed by combining the Q and U images with the {\it miriad} task {\sc IMPOL}, and were later rotated by $\frac{\pi}{2}$ to convert them into magnetic vectors. The observational curve has two maxima: a large one at
 $\sim 57\degr$ and a small one at $\sim25\degr$. Figure
 \ref{gaussobs} shows a fitting of the observational distribution by
 two Gaussian labeled as g1 and g2, whose parameters are given in
 Table 2. The mean value of Gaussian g1 is 58\degr, which is close to
 the inclination of the Galactic plane (60\degr) and in agreement with
 the maxima of the curves computed for the runs R1a and R2a (see Figure
 \ref{compnonoise}).

It is reasonable to expect that the resulting ambient magnetic field is parallel to the Galactic plane at the latitude of SN 1006
\citep{bocchino2011,reynoso2013,schneiter2015}. The secondary peak of
the observed position-angle distribution would indicate an extra
magnetic field component, which is not considered in our
simulations. 
In order to eliminate this secondary unmodelled component, we subtracted the Gaussian fit labeled g2 from the observed distribution. The resulting curve will be referred to as the modified
observational distribution.

Figure \ref{compnoise} compares the modified observational
distribution with those resulting from the runs R1b and R2b. Clearly these
numerical curves display now wider distributions compared with runs R1a and R2a, which is a consequence of
including perturbations in {\bf B} and $\rho$. The maxima of
numerical distributions remain close to 60\degr .

\begin{figure}
  \centering
   \includegraphics[width=0.5\textwidth]{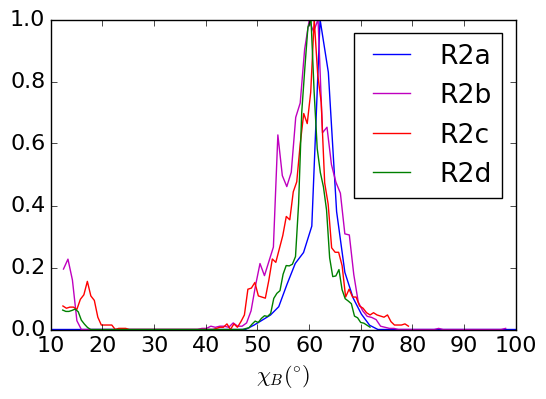}
    \caption{Comparison of the position-angle distributions obtained
      for runs R2p, i.e. those considering  $\gamma=1.3$. 
       In model R2a the medium is initialized as uniform at the beginning of the simulation, while in model R2b the SNR propagates into a turbulent medium (i.e., $\delta B, \delta \rho \neq 0$. In models R2c and R2d the medium is ``partially'' turbulent, with $\delta B\neq 0, \delta \rho = 0$ and $\delta \rho\neq 0, \delta B = 0$ respectively.
}%
\label{compg13}
\end{figure}

In Figure \ref{compg13} we explore the effect of selectively including the perturbations in $B$ and/or $\rho$ for $\gamma$ = 1.3. The
curve corresponding to the run R2d ($\delta \rho\,\neq\,0$ and $\delta${\bf
  B}$\,=0$) does not show any significant widening, in contrast with those
corresponding to the runs R2b and R2c, both with $\delta${\bf B}$\,\neq 0$. 
In order to carry out a more quantitative study, a statistical
analysis was performed, comparing the distributions of the position-angle obtained from observations (subtracting the contribution of Gaussian
g2) and simulations. The results of this statistical analysis are
summarized in Table 3. Comparing the values obtained for the standard
deviation, skewness and kurtosis, we note that the models achieving a better fit with the observations are those that consider a perturbed magnetic field, and a lower $\gamma$ (runs R2b and R2c).

\begin{table}
\begin{center}
  \begin{tabular}{lrrrr}
            \hline
            \noalign{\smallskip}
     \ \   &  mean($\degr$) &  $\sigma$($\degr$) & skewness & kurtosis\\
             \hline
           \noalign{\smallskip}
R1a  &  $60.$  &  $3.5$ & $-0.32$ & $0.2$\\
R1b  &  $59.$  &  $7.5$ & $0.52$ & $5.2$\\
R2a  & $62.$  &  $6.6$ & $-5.80$ & $91.0$\\
R2b  &  $58.$  &  $11.0$ & $-3.0$ & $9.7$\\
R2c  &  $56.$  &  $13.0$ & $-2.2$ & $4.7$\\
R2d  &  $58.$  &  $9.9$ & $-3.70$ & $14.0$\\
obs-g2  &  $57.$  &  $22.0$ & $-0.68$ & $6.5$\\
           \hline
            \noalign{\smallskip}
  \end{tabular}
\caption{Statistics of the position-angle distributions}
\end{center}

\label{tab3}
\end{table}

\section{Discussion and conclusions}\label{discussion}

A radio polarization study of SN 1006 was
  performed based on 3D MHD simulations, considering the expansion of the remnant into a turbulent interstellar medium (including both magnetic field and density perturbations).

The inclusion of perturbations in magnetic field and
  density do not change the main conclusion of
  \citet{schneiter2015},  namely that the quasi-parallel case is the acceleration
  mechanism that better explains the observed morphology 
in the distribution of the
  Stokes parameter Q. 

Based on the polar-referenced angle method, a study of the position angle distribution of the magnetic field was
  performed on both the observations and numerical results. This study
  reveals that the observational distribution is wide and has two
  maxima or components: a large one at 58\degr that coincides
  with the expected direction of the ambient magnetic field, and a
  small component at 23\degr. The curves corresponding to runs
  R1a an R2a have a single peak and narrower distributions. The  secondary peak of the observations could be due to local blow-outs produced during the expansion of parts of the main SNR shock front into low-density cavities, such as  those explored by \citet{yu2015}. Since our simulations do not have this secondary component we filtered it our from the observations and focused our comparison with the dominant component, as explained in the previous section.

The subsequent analysis was carried out by comparing
  the position-angle distributions obtained from numerical simulations
  with the observational distribution curve. A statistical
  study performed on these distributions reveals that
  models which include
  a turbulent pre-shock magnetic field, successfully
  explain the observed polar-angle distribution. Furthermore, all of
  them share a maximum at a position angle of 60$\degr$, which is parallel to the Galactic Plane and coincides with the expected direction of the Galactic magnetic field around
  SN1006 \citep{bocchino2011,reynoso2013}. 

  In summary, the statistical analysis based on the polar-referenced
  angle technique carried out in this work made it possible to recover
  the orientation of the ambient magnetic field. Based on
    previous observational and theoretical studies of SN1006 \citep{bocchino2011,reynoso2013,schneiter2015}, we have only considered
    the quasi-parallel acceleration mechanism to estimate the
    synchrotron emission. However, it is important to mention that the
    same technique presented here can be applied with the
    quasi-perpendicular case.
  Finally, a good agreement between observations and numerical results
  is obtained if the simulations include magnetic field perturbations.

\section*{Acknowledgments}

The authors thank the referee for reading this manuscript and for
her/his useful comments. PFV, AE, and FDC acknowledge finantial support from DGAPA-PAPIIT (UNAM) grants IG100516, IN109715, IA103315, and
IN117917.  EMS, EMR, and DOG are members of the Carrera del
Investigador Cient\'\i fico of CONICET, Argentina. MVS and AM-B are
fellows of CONICET (Argentina) and CONACyT (M\'exico),
respectively. EMS thanks Paulina Vel\'azquez for her hospitality during
his visit to Mexico City. We thank Enrique Palacios-Boneta (c\'omputo-ICN) for mantaining the Linux servers where our simulations were carried out.

\bibliography{ref}{}
\bibliographystyle{mnras}

\bsp

\label{lastpage}

\end{document}